\documentclass[10pt,twocolumn]{article} 
\usepackage{simpleConference}
\usepackage{spconf,amsmath,amssymb,graphicx,booktabs} 
\usepackage[ruled,linesnumbered]{algorithm2e} 
\usepackage{url, hyperref}
\begin{document}

\title{Geometrically Matched Multi-source Microscopic Image Synthesis \\ Using Bidirectional Adversarial Networks}

\twoauthors{Jun Zhuang}
{Indiana University-Purdue University Indianapolis\\
Indianapolis, IN 46202 \\
junz@iu.edu}
{Dali Wang \thanks{This study is supported by an NIH research project grants (R01GM097576). }}
{University of Tennessee\\
Knoxville, TN 37996 \\
dwang7@utk.edu}

\maketitle
\thispagestyle{empty}

\begin{abstract}
Microscopic images from multiple modalities can produce plentiful experimental information. In practice, biological or physical constraints under a given observation period may prevent researchers from acquiring enough microscopic scanning. Recent studies demonstrate that image synthesis is one of the popular approaches to release such constraints. Nonetheless, most existing synthesis approaches only translate images from the source domain to the target domain without solid geometric associations. To embrace this challenge, we propose an innovative model architecture, BANIS, to synthesize diversified microscopic images from multi-source domains with distinct geometric features. The experimental outcomes indicate that BANIS successfully synthesizes favorable image pairs on {\em C. elegans} microscopy embryonic images. To the best of our knowledge, BANIS is the first application to synthesize microscopic images that associate distinct spatial geometric features from multi-source domains.
\end{abstract}


\section{Introduction}
\label{sec:intro}
\vspace{-3mm}
Multi-source observation, which observes the same objective from different sources, has been widely used in many different areas, such as biology and medical fields \cite{su2020medical, zhang2019high, mou2020cs2, meinel2007breast, yuan2019automated, liu2019granular, femmam2015optimizing, lu2015automated, huo2018synseg, zhao2015automated, peng2016determination, dai2012aqueous, nie2018medical, zamzmi2020accelerating, vakharia2019effect, parisi2020supervised, wei2019precise, gao2016corner}. For example, microscopic imaging of cell nucleus and membrane separately, with different fluorescent materials, is one kind of multi-source observations. 

Cross-domain synthesis \cite{dar2019image, miller1993mathematical, lee2017multi, jog2015mr, chartsias2017multimodal} is one potential solution to augment multi-source observation. Given a source domain \textit{A}, cross-domain synthesis aims at generating corresponding images of the same objective in a target domain \textit{B}, or vice versa.
According to \cite{dar2019image}, such synthesis can be divided into two main types, the registration-based \cite{miller1993mathematical, lee2017multi} and the intensity-transformation-based methods \cite{jog2015mr, chartsias2017multimodal}. The registration-based method assumes that images within both the source domain and the target domain are geometrically associated with each other. This method generates images from a co-registered set of images \cite{miller1993mathematical}. On the other hand, the intensity-transformation-based method does not fully rely on the geometric relationship. For example, multimodal is a deep learning approach for MRI image synthesis \cite{chartsias2017multimodal}. The model takes multi-source images as input from source contrasts and yields high-quality images in the target contrast. However, both types of methods mentioned above could not solve the issue that two domains come from different sources with quite different spatial features. 

In this study, we propose a novel model, \textbf{B}idirectional \textbf{A}dversarial \textbf{N}etworks for microscopic \textbf{I}mage \textbf{S}ynthesis (\textbf{BANIS}), which uses bidirectional adversarial network to synthesize geometrically matched images from multiple domains. BANIS, to the best of our knowledge, is the first cross-domain synthesis application with multi-source images of entirely separated spatial patterns. In the experiment, we deploy our model to a set of microscopic images from {\em C. elegans} embryogenesis. The experimental results demonstrate that BANIS successfully synthesizes diversified, geometrically matched microscopic images and outperforms two baseline models.
Our contribution in this work can be summarized as follows:
\begin{description}
  \item[$\bullet$\ ] We propose a novel model, BANIS, to synthesize geometrically matched images from multiple domains. To the best of our knowledge, BANIS is the first cross-domain synthesis application with multi-source images of entirely separated spatial patterns.
  \item[$\bullet$\ ] The experiments indicate that BANIS successfully synthesizes diversified, geometrically matched microscopic images and outperforms two baseline models. We will make the model source code and the {\em C. elegans} embryo microscopic dataset publicly available after the paper acceptance \footnote{Our code is available on Github at: \url{https://github.com/junzhuang-code/BANIS}}.
\end{description}

\section{Methodology}
\label{sec:method}
\vspace{-2mm}
\subsection{Preliminary Background}
\vspace{-3mm}
The Generative Adversarial Network (GAN) is one of popular deep learning techniques for cross-domain synthesis \cite{goodfellow2014generative, donahue2016adversarial}. Vanilla GAN \cite{goodfellow2014generative} consists of two key components, a generator $\textit{G}$ and a discriminator $\textit{D}$. Given a prior distribution $Z$ as input, $\textit{G}$ maps a point $\mathbf{z} \sim Z$ from the latent space to the data space as $\textit{G}(\mathbf{z})$. On the other hand, $\textit{D}$ attempts to distinguish an instance $\mathbf{x}$ from a synthetic instance $\textit{G}(\mathbf{z})$, generated by $\textit{G}$. The training process is set up as if $\textit{G}$ and $\textit{D}$ are playing a zero-sum game. On the one hand, $\textit{G}$ tries to generate the synthetic instances that are as close as possible to real instances. On the other hand, $\textit{D}$ distinguishes the synthetic instances from the real instances. After the model converges, both $\textit{G}$ and $\textit{D}$ reach a Nash equilibrium. At this point, $\textit{G}$ is able to generate instances which are very close to the real one. The objective function $\textit{V}$ of Vanilla GAN can be written as a summation of two Expectation values $\mathbb{E}$ as follows:
\begin{equation}
\begin{split} 
\min_{\textit{G}} \max_{\textit{D}} V(\textit{D}, \textit{G}) = &\underset{\mathbf{x} \sim {X}} {\mathbb{E}} [\log \textit{D}(\mathbf{x})] + \\ 
&\underset{\mathbf{z} \sim {Z}} {\mathbb{E}} [\log (1-\textit{D}(\textit{G}(\mathbf{z})))]
\end{split}
\end{equation}
where $X$ and $Z$ are the corresponding distribution that $\mathbf{x}$ and $\mathbf{z}$ are sampled from.

In cross-domain synthesis, however, many instances are unpaired between domains \cite{zhu2017unpaired}. Zhu et al. \cite{zhu2017unpaired} propose cycle-consistent loss to map the synthetic instances as close as possible to the original instances through the cycled generation, which is combined with two sets of generators and discriminators. The cycle-consistent loss function is described as follows:
\begin{equation}
\begin{split} 
\textit{L}_{cyc} (\textit{G}_{A}, \textit{G}_{B}) = 
&\underset{\mathbf{a} \sim {A}} {\mathbb{E}} [\| \textit{G}_{B}(\textit{G}_{A}(\mathbf{a})) - \mathbf{a} \|] + \\
&\underset{\mathbf{b} \sim {B}} {\mathbb{E}} [\| \textit{G}_{A}(\textit{G}_{B}(\mathbf{b})) - \mathbf{b} \|]
\end{split}
\end{equation}
where $\mathbf{a}$ and $\mathbf{b}$ are instances from domain $A$ and $B$.

\subsection{Model Architecture}
\begin{figure}[h]
  \centering
  \includegraphics[width=\linewidth]{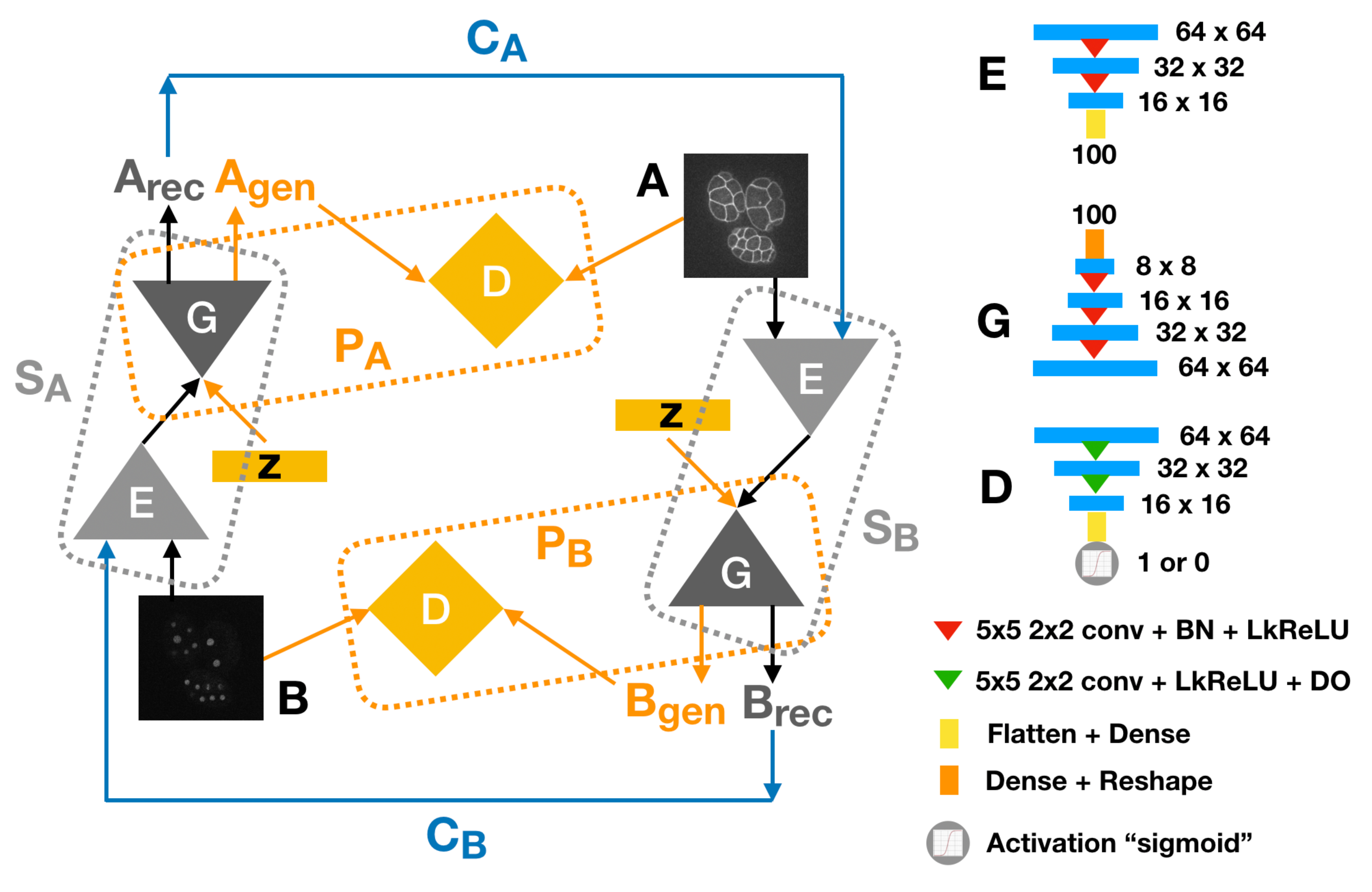}
  \caption{Our model, BANIS, contains two Pioneers, two Successors, and two Coordinators. The Pioneer is composed of a generator $\textit{G}$ and a discriminator $\textit{D}$. With random uniform priors $\mathbf{z}$ as inputs, $\textit{P}_{A}$ and $\textit{P}_{B}$ generate images $A_{gen}$ or $B_{gen}$, respectively. The Successor consists of an encoder $\textit{E}$ and shares the generator $\textit{G}$ with the Pioneer. $\textit{S}_{A}$ and $\textit{S}_{B}$ learn the prior knowledge from observed images, $B$ and $A$, and reconstruct new images $A_{rec}$ or $B_{rec}$. By sequentially connecting two Successors, the Coordinators $\textit{C}_{A}$ and $\textit{C}_{B}$ are designed  to reinforce the spatial similarity between the reconstructed images, $B_{rec}$ and $A_{rec}$, and the observed images $B$ and $A$, separately.
  The right side of Figure \ref{fig:model} shows an exemplar architecture of $\textit{E}$, $\textit{G}$ and $\textit{D}$. The number indicates the size of network layer. For example, encoder $\textit{E}$ takes 64 $\times$ 64 image as input and outputs a 100-dimension vector of latent variables. $5\times5 \ 2\times2 \ conv$ represents 2D convolutional layer with $5 \times 5$ kernel size and $2 \times 2$ strides. $BN$, $LkReLU$ and $DO$ stand for batch normalization layer, LeakyReLU activation layer, and dropout layer, respectively.}
\label{fig:model}
\end{figure}

\vspace{-3mm}
In this paper, we propose a novel model, Bidirectional Adversarial Networks for microscopic Image Synthesis (BANIS). As displayed in Figure \ref{fig:model}, BANIS contains two Pioneers $\textit{P}$, two Successors $\textit{S}$ and two Coordinators $\textit{C}$. The Pioneer is composed of a Generator $\textit{G}$ and a Discriminator $\textit{D}$. The $\textit{P}$ is mainly responsible for pre-training in the warm-up stage to speed up the progress of synthesis.
The Successor consists of an Encoder $\textit{E}$ and shares the Generator $\textit{G}$ with the Pioneer. The $\textit{S}$ uses its $\textit{E}$ to compress an input image into latent variables and then uses its $\textit{G}$ to reconstruct the new image from these latent variables. 
The Coordinator uses pixel-wise methods to preserve the geometric relationship between the images reconstructed by two Successors and the original observed images. 

The training procedure of BANIS contains two stages. BANIS simultaneously takes the input images from domain $A$ and domain $B$. On the warm-up stage, $\textit{P}_{A}$ and $\textit{P}_{B}$ are trained with given random uniform priors $\mathbf{z}$ and then respectively generate images $A_{gen}$ and $B_{gen}$. Preliminary images start forming without geometric matching between these images. After the warm-up stage, both $\textit{S}$ and $\textit{C}$ join the training to enforce the geometrical relationship between these synthesized images. $\textit{S}_{A}$ and $\textit{S}_{B}$ learn prior knowledge from observed images, $B$ and $A$, and reconstruct images with its pre-trained generators from the Pioneers. At the same time, $\textit{C}_{A}$ and $\textit{C}_{B}$ respectively reinforce the spatial similarity between observed images, $B$ or $A$, and reconstructed images, $B_{rec}$ and $A_{rec}$, separately. The model does not stop training until the geometric relationship between image pairs forms. After that, we decrease the learning rate of both $\textit{S}$ and $\textit{C}$ on subsequent training to improve the quality of synthesized images.

\subsection{Loss Functions}
\vspace{-3mm}
BANIS uses three types of loss functions, Adversarial Loss, Identical Loss, and Pair-matched Loss, to help synthesize geometrically matched images.

\textbf{Adversarial Loss} \cite{goodfellow2014generative} is employed to enforce the generated image $A_{gen}$ or $B_{gen}$as similar as possible to the observed image $A$ or $B$. The adversarial loss applies to the Pioneer in the whole training process. Given a random uniform prior, however, generated images don’t preserve the spatial information between multiple source domains. Note that our model synthesizes the pair of images simultaneously. Thus, this loss applies to both domain $A$ and domain $B$. Here we use the same denotation as Vanilla GAN.
\begin{equation}
\begin{split} 
\textit{L}_{adv} (\textit{G}, \textit{D}) = &\underset{\mathbf{x} \sim {X}} {\mathbb{E}} [\log \textit{D}(\mathbf{x})] + \\ 
&\underset{\mathbf{z} \sim {Z}} {\mathbb{E}} [\log (1-\textit{D}(\textit{G}(\mathbf{z})))]
\end{split}
\label{eq:adv}
\end{equation}

\textbf{Identical Loss} applies to the Successor. To solve previous limitations, the Successor takes specific prior and attempts to reconstruct the images $A_{rec}$ or $B_{rec}$. Note that the Pioneer helps speed up the synthesis in the warm-up stage. Pioneer’s generator is shared with Successor. In other words, reconstructed images are expected to be as close as possible to both observed images $A$ or $B$ and generated images $A_{gen}$ or $B_{gen}$. Identical loss ensures the quality of reconstructed images. In this paper, we use mean squared error (MSE) to measure the similarity.
\begin{equation}
\begin{split} 
\textit{L}_{id} (\textit{S}_{A}, \textit{S}_{B}, \textit{G}_{A}, \textit{G}_{B}) =
&\underset{\mathbf{b} \sim {B}, \mathbf{a} \sim {A}} {\mathbb{E}} [\| \textit{S}_{A}(\mathbf{b}) - \mathbf{a} \|] + \\
&\underset{\mathbf{b} \sim {B}, \mathbf{z} \sim {Z}} {\mathbb{E}} [\| \textit{S}_{A}(\mathbf{b}) - \textit{G}_{A}(\mathbf{z}) \|] \\
&\underset{\mathbf{a} \sim {A}, \mathbf{b} \sim {B}} {\mathbb{E}} [\| \textit{S}_{B}(\mathbf{a}) - \mathbf{b} \|] + \\
&\underset{\mathbf{a} \sim {A}, \mathbf{z} \sim {Z}} {\mathbb{E}} [\| \textit{S}_{B}(\mathbf{a}) - \textit{G}_{B}(\mathbf{z}) \|]
\end{split}
\label{eq:id}
\end{equation}

\textbf{Pair-matched Loss} applies to the Coordinator. This loss enforces the projection inside each Successor to ensure these two domains are spatially matched. The Coordinator sequentially connects these two Successors. $\textit{C}_{A}$ takes $B$ as input and uses $\textit{S}_{A}$ and $\textit{S}_{B}$ sequentially to generate $B_{rec}$. Then it compares the new images with the observed image $B$. $\textit{C}_{B}$ operates similarly with observed image $A$. In other words, pair-matched loss helps preserve spatial information among two domains. In this paper, we also use MSE to measure the quality of the projection.
\begin{equation}
\begin{split} 
\textit{L}_{pm} (\textit{C}_{A}, \textit{C}_{B}) = &\underset{\mathbf{b} \sim {B}} {\mathbb{E}} [\| \textit{C}_{A}(\mathbf{b}) - \mathbf{b} \|] + \\
&\underset{\mathbf{a} \sim {A}} {\mathbb{E}} [\| \textit{C}_{B}(\mathbf{a}) - \mathbf{a} \|]
\end{split}
\label{eq:pm}
\end{equation}

\begin{algorithm}
\DontPrintSemicolon 
\KwIn{Testing set $(A_{test}, B_{test})$, Threshold $TS$}
Initialize two counters, $cnt_{total}$ and $cnt_{matched}$ as 0; \\
\For{$\textbf{all} \ (a^{\{i\}}, b^{\{i\}}) \in (A_{test}, B_{test})$} {
  $cnt_{total} \gets cnt_{total} + 1;$ \\
  $a^{\{i\}}_{rec}, b^{\{i\}}_{rec} = \textit{S}_{B}(a^{\{i\}}), \textit{S}_{A}(b^{\{i\}});$ \\
  $a^{\{i\}}_{bi}, b^{\{i\}}_{bi} = \textit{Bi}(a^{\{i\}}_{rec}), \textit{Bi}(b^{\{i\}}_{rec});$ \\
  $Dice_{AB} = \textit{DSC}(a^{\{i\}}_{bi}, b^{\{i\}}_{bi});$ \\
  \If{$Dice_{AB} < TS$} {
    $cnt_{matched} \gets cnt_{matched} + 1;$ \\
  }
}
\Return{$\frac{cnt_{matched}}{cnt_{total}}.$}
\caption{{\sc Geometric} Matching Index (GMI)}
\label{algo:gmi}
\end{algorithm}

\subsection{Geometric Matching Index}
\vspace{-3mm}
Although the individual synthesized images come from different domains and have different geometric patterns, the pair of images should be geometrically matched. For example, in our microscopic data case, membranes and nucleus should be spatially matched without overlapping. For this purpose, we propose a new evaluation metric, Geometric Matching Index (GMI), to measure the quality of image synthesis. As the pseudo-code presented in Algorithm \ref{algo:gmi}, GMI first extracts the binary masks from reconstructed image pairs and then measures their contours' overlapping by the Dice Similarity Coefficient (DSC) \cite{zhuang2019lighter}. Less overlapping in our case means better matching. Given an overlapping threshold, GMI counts the number of well-matched image pairs whose DSC is lower than the given threshold and returns the percentage of these well-matched image pairs over the total number of reconstructed image pairs at the end.

\begin{figure}[h]
  \centering
  \includegraphics[width=\linewidth]{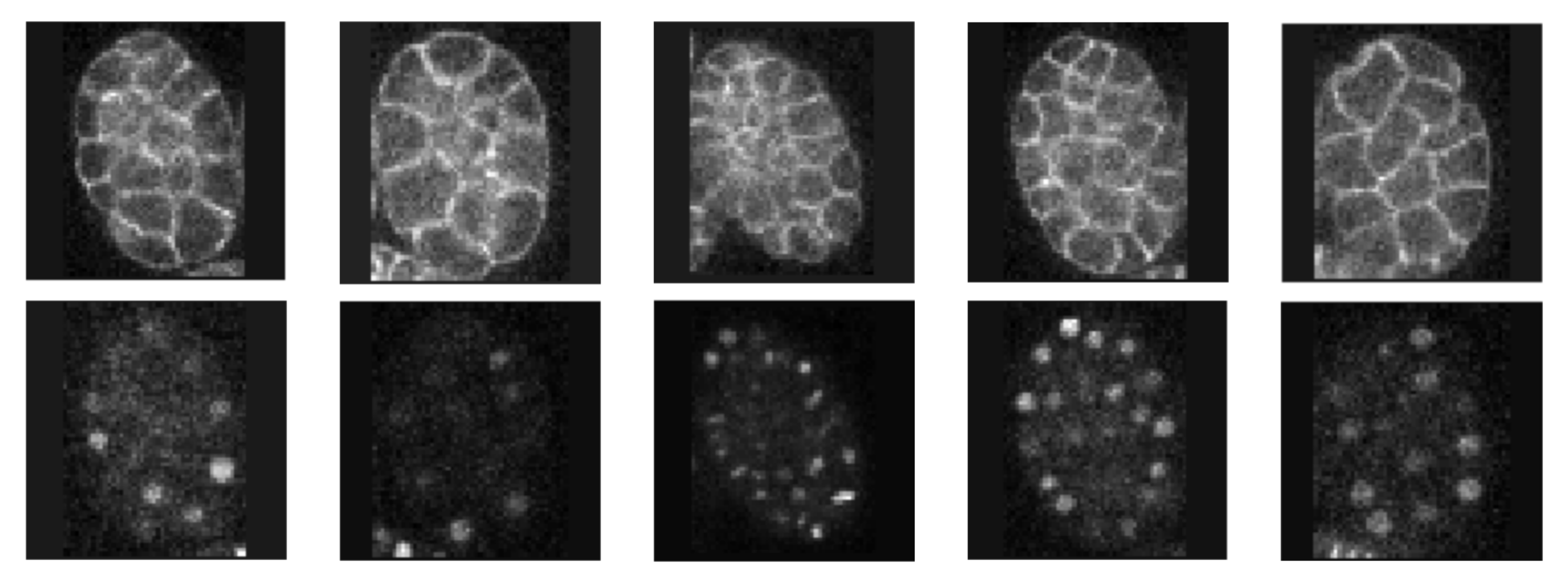}
  \caption{Samples of the Observed Microscopic Images}
\label{fig:img_obs}
\end{figure}

\section{Experiments}
\label{sec:exp}
\vspace{-2mm}
\subsection{Dataset and Preprocessing}
\vspace{-3mm}
In this experiment, our model is evaluated on a set of {\em C. elegans} microscopy image dataset \cite{wang2019}. Each set contains 300 voxels, and each of them may contain one to three embryos. The scanner takes 75-second intervals on each voxel over the first 375 minutes of embryogenesis. We select 50 pseudo-3D voxels and each of them contains three embryos. Each voxel contains 30 slices at 1 $\mu m$ vertical distance that covers the entire embryo(s). Each slice contains one 512 $\times$ 512 membrane image and one 512 $\times$ 512 nuclei image.

We use ImageJ to split these image stacks and pick the raw images from the middle 15 layers of each stake for our model experiments. We split the raw images into two 512 $\times$ 512 images, each contains membrane- or nuclei-only information. These 512 $\times$ 512 images are then converted to gray scale and denoised with a Gaussian filter. Then, we resized these images into 128 $\times$ 128 for a better computational efficiency. After that, we crop out a single embryo and generate smaller 64 $\times$ 64 images. Finally, we normalize the pixel value of the image between -1 and 1. 10\% of images is used as a test set and the rest part is for training. Some samples of the 64x64 microscopic images are illustrated in Figure \ref{fig:img_obs}.

\subsection{Model Parameters and Training}
\vspace{-3mm}
The shape of input images is $64 \times 64 \times 1$. The latent dimension for both random uniform prior and encoded prior is 100. Both $\textit{D}$ and $\textit{P}$ are trained with Adam optimizer. The initial learning rate for $\textit{G}_{B}$ is set as $1 \times 10^{-5}$ and the learning rates for $\textit{G}_{A}$, $\textit{D}_{A}$ and $\textit{D}_{B}$ are set as $2 \times 10^{-5}$. Both $\textit{S}$ and $\textit{C}$ are trained by Stochastic Gradient Descent (SGD) optimizer. Their initial learning rates are $1 \times 10^{-4}$. The batch size is empirically set as 128.

The model is trained with 17,000 epochs in the warm-up stage at the initial learning rate until adversarial the loss converges and the preliminary images start forming. After that, both $\textit{S}$ and $\textit{C}$ join the training for subsequent 13,000 epochs until both identical loss and pair-matched loss converge. To improve the quality of synthesized images, both $\textit{S}$ and $\textit{C}$ are trained with another 10,000 epochs at a new learning rate, which is decreased by 50\%. The total model training with 40,000 epochs takes approximately 12.3 hours on a Linux machine, which configured with 4 Intel Xeon central processing units (E5-1620 v4), 64-GB memory, and a 16-GB Nvidia GP104 graphical processing unit.

After the training, we evaluate the BANIS performance by calculating the GMIs of the entire synthesized dataset with different overlapping thresholds $TS_{dsc}$. A lower $TS_{dsc}$ indicates a more strictly geometric matched is required.

\subsection{Experimental Results}
\begin{figure}[h]
  \centering
  \includegraphics[width=\linewidth]{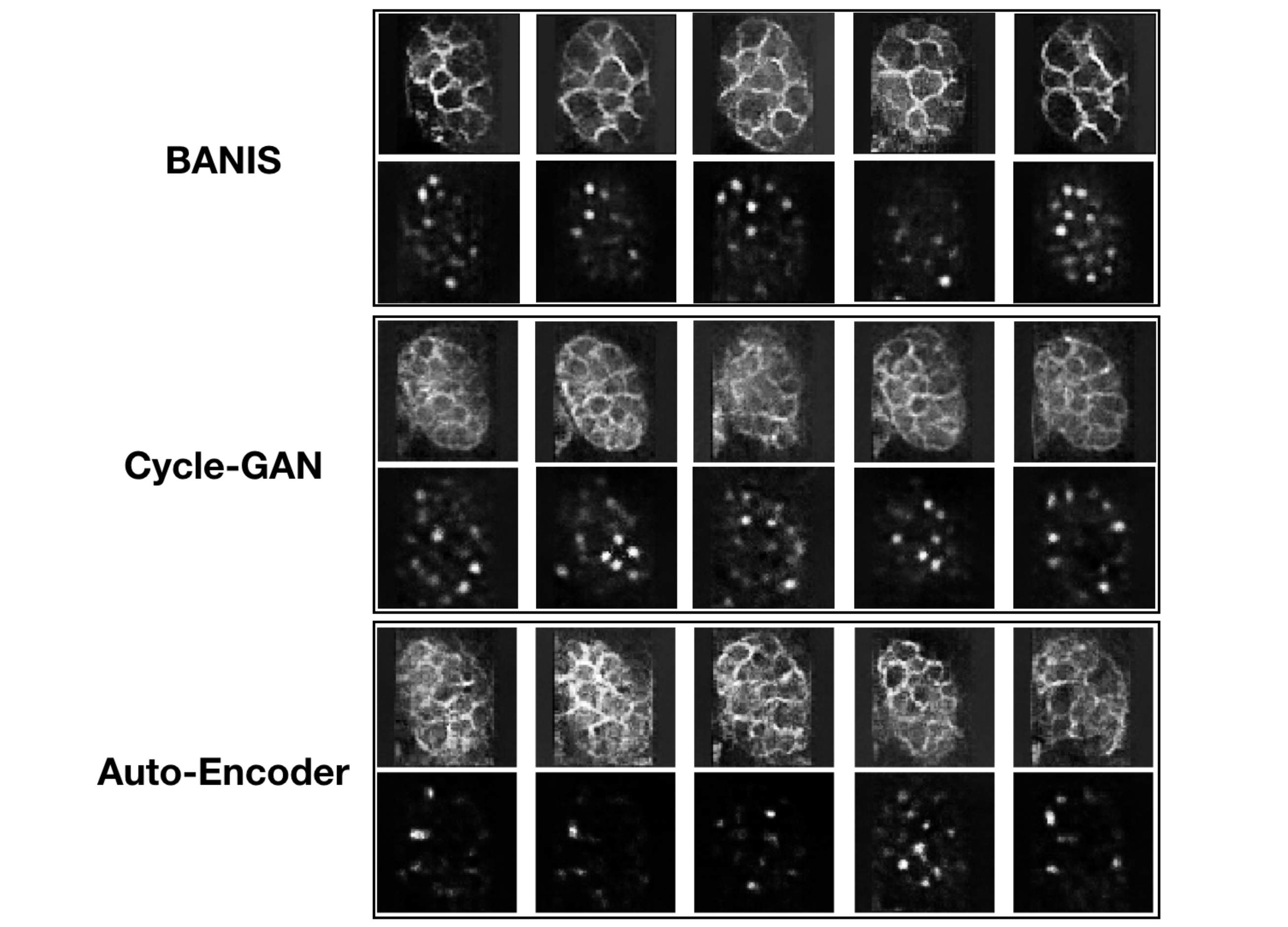}
  \caption{Exemplar Synthesized Images From BANIS, Cycle-GAN, And Auto-Encoder}
\label{fig:img_match}
\end{figure}

\vspace{-3mm}
In this experiment, we first examine the performance of synthesis between BANIS and baselines. We select Cycle-GAN \cite{zhu2017unpaired} and Auto-Encoder \cite{kingma2013auto} as our baselines since BANIS is inspired by both of them. We train Cycle-GAN / Auto-Encoder to converge with 200 / 300 epochs, respectively. Rest experimental settings remain the same as BANIS. Figure \ref{fig:img_match} presents the synthesized images. BANIS can synthesize geometrically matched image pairs (the 1st and 2nd rows). These synthesized images have clear {\em C. elegan's} image features of membrane or nuclei and simultaneously preserve the geometric relationship between them. It is clear that these synthesized image pairs have very similar patterns shown in Figure \ref{fig:img_obs}. On the contrary, the synthesized images from Cycle-GAN (the 3rd and 4th rows) couldn't preserve the geometric relationship as Cycle-GAN is only good at transferring the style or texture between two images with similar shapes. Auto-Encoder only generates fuzzy images with unclear contours (the 5th and 6th rows). What's worse, the nuclei are barely seen in some synthesized samples. We argue that partial information irreversibly gets loss in the encoding stage, which leads to this unsatisfied synthesis. Note that these images are slices and thus may not display all nuclei in one slice. These synthesized images demonstrate that BANIS achieves superior performance against two baselines.

\begin{table}[h]
\small
\centering
\setlength{\tabcolsep}{3pt}
  \caption{Evaluation On The Synthesized Images By GMI}
  \vspace{3mm}
  \label{tab:smi}
  \begin{tabular}{ccccc}
    \toprule
    \textbf{$TS_{dsc}$ (\%)} & 0.1 & 0.2 & 0.3 \\
    \midrule
    \textbf{BANIS} & 75.26 ($\pm$ 0.65) & 87.27 ($\pm$ 0.32) & 95.39 ($\pm$ 0.11) \\
    \textbf{Cycle-GAN} & 69.04 ($\pm$ 0.48) & 82.89 ($\pm$ 0.67) & 90.14 ($\pm$ 0.28) \\
    \textbf{Auto-Encoder} & 71.48 ($\pm$ 1.21) & 83.86 ($\pm$ 0.98) & 90.91 ($\pm$ 0.51) \\
  \bottomrule
\end{tabular}
\end{table}

We also evaluate the performance based on the aforementioned metric, GMI. We run each experiment five times and present the mean and standard deviation in Table \ref{tab:smi}. The outcome reveals that BANIS yields satisfied synthesized images under strict thresholds $TS_{dsc}$ and outperforms the other two baselines across three different thresholds. Most (over 95\%) images of the total image pairs have an overlapping value less than 0.3. Even with a very restricted threshold requirement of 0.1 (that is less than 10\% of overlapping between any two simultaneously synthesized images), the GMI of the total synthesized image pairs reaches 75.26\%. We observe that Auto-Encoder achieves higher GMI, but it fails to synthesize nuclei images. We argue that this failure decreases the overlapping and thus increases GMI. Overall, GMI is a qualified metric to measure the geometrical matching relationship between two synthesized domains. However, it sometimes fails if other reasons weaken the overlapping as well.

\section{Conclusion}
\label{sec:con}
\vspace{-3mm}
In this study, we present an innovative model architecture, BANIS, to synthesize microscopic images from multiple domains. BANIS, to the best of our knowledge, is the first model that synthesizes geometrically matched images from multiple domains that exist entirely separated spatial patterns. The experiment using microscopic data from {\em C. elegan's} embryogenesis proves that our model can synthesize diversified and geometrically matched images that are as comparable as the observed microscopic images.

\bibliographystyle{abbrv}
\bibliography{reference}
\end{document}